\documentclass[showpacs,aps,superscriptaddress,prl,preprint, amsmath]{revtex4}
\usepackage{graphicx,subfigure,float}
\usepackage[percent]{overpic}
\usepackage{amsmath}
\usepackage{amssymb}
\usepackage[colorlinks, citecolor=blue]{hyperref}
\usepackage{subfigure}
\begin{document}

\preprint{APS/123-QED}

\title{Second-Order Topological Insulator in Two-Dimensional C$_{2}$N and Its Derivatives}
\author{ZeHou Li}
\affiliation{Hunan Provincial Key laboratory of Thin Film Materials and Devices, School of Material Sciences and Engineering, Xiangtan University, Xiangtan 411105, People's Republic of China}
 \affiliation{Hunan Key Laboratory for Micro-Nano Energy Materials and Devices, School of Physics and Optoelectronics,
Xiangtan University, Hunan 411105, People's Republic of China}
 \author{Pan Zhou}
 \email{zhoupan71234@xtu.edu.cn}
 \affiliation{Key Laboratory of Low-dimensional Materials and Application Technology, School of Material Sciences and Engineering, Xiangtan University, Xiangtan 411105, People's Republic of China}
\author{QiuHui Yan}
 \affiliation{Hunan Provincial Key laboratory of Thin Film Materials and Devices, School of Material Sciences and Engineering, Xiangtan University, Xiangtan 411105, People's Republic of China}
  \author{XiangYang Peng}
  \email{xiangyang\_peng@xtu.edu.cn}
 \affiliation{Hunan Key Laboratory for Micro-Nano Energy Materials and Devices, School of Physics and Optoelectronics,
Xiangtan University, Hunan 411105, People's Republic of China}
 \author{ZengSheng Ma}
 \affiliation{Hunan Provincial Key laboratory of Thin Film Materials and Devices, School of Material Sciences and Engineering, Xiangtan University, Xiangtan 411105, People's Republic of China}
\author{LiZhong Sun}
 \email{lzsun@xtu.edu.cn}
 \affiliation{Hunan Provincial Key laboratory of Thin Film Materials and Devices, School of Material Sciences and Engineering, Xiangtan University, Xiangtan 411105, People's Republic of China}
\date{\today}
\begin{abstract}
\indent High-order topological phase exhibits nontrivial gapless states at the boundaries whose dimension is lower than bulk by two. However, this phase has not been observed experimentally in two-dimensional (2D) materials up to now. In this work, using first-principles calculations and tight-binding (TB) model, we propose that the experimentally synthesized C$_{2}$N is a 2D second-order topological insulator(SOTI) with one-dimensional gapped edge states and zero-dimensional gapless corner states. C$_2$N exhibits a large bulk gap of 2.45 eV and an edge gap of 0.32 eV, making it an excellent candidate to evidently present the nontrivial corner states in experiments. The robustness of the corner states against the edge disorders has been explicitly identified. Moreover, another three C$_2$N-like materials are also found to host the nontrivial SOTI phase including an experimentally synthesized material aza-fused $\pi$-conjugated microporous polymers (aza-CMP). The four 2D SOTIs proposed in our present work provide excellent candidates for studying the novel high-order topological properties in future experiments.\\
\end{abstract}
 \maketitle
\section*{INTRODUCTION}
\indent In recent years, topological insulators have attracted enormous attentions due to their intrinsic topological nature and potential applications in many fascinating areas, such as low-power electronic devices\cite{lpe-1,lpe-2,lpe-3,lpe-4} and quantum computing\cite{TI-QC1,TI-QC2}. A conventional $d$-dimensional topological insulator supports gapless states on its ($d$-1)-dimensional boundaries, and the states are insensitive to the local perturbations that preserve the nontrivial topology of the bulk\cite{TI1,TI2,TI3}. Recently, the notion of second-order topological insulators (SOTIs) has been proposed to extend the bulk-boundary correspondence, in which the gapless states appear in the corners of a 2D system or the hinges of a three-dimensional system\cite{hoti-1,hoti-2,hoti-3,hoti-4,hoti-5,hoti-6,hoti-7,hoti-8,hoti-9,hoti-10,hoti-11,kekule1,kagome}. From a physical point of view, the low-dimensional topological boundary states are determined by electric dipole or quadrupole moments\cite{hoti-1,hoti-2}. So far, the topological corner states were mainly observed in artificial 2D systems\cite{arti-sys-1,arti-sys-2,arti-sys-3,arti-sys-4,arti-sys-5,addition1}, such as square, Kagome, or hexagonal lattices, in which the nontrivial phases are mainly derived from the difference between intercell and intracell hopping parameters. However, the hoppings in real electronic materials are generally more complicated. Therefore, we need to go beyond the simplified model to find more potential SOTIs. Although some experimentally synthesized materials, such as monolayer graphdiyne\cite{GDY-1,GDY-2}, $\gamma$-graphyne\cite{GY}, large-angle twisted bilayer graphene\cite{bilayer}, honeycomb antimony\cite{topo_mat1}, and twisted bilayer graphene and twisted bilayer boron nitride\cite{TBBN}, have been predicted as SOTIs in their low-energy gaps, none of them has been experimentally verified yet. The main obstacle for the confirmation of a SOTI in a realistic system is that its bulk and edge states gaps should be large enough so that the corner states can be isolated from the bulk and edge states and not be destroyed by external perturbations. Therefore, finding SOTIs fulfilling above requirements is an urgent task in the area.\\
\indent Recently, a 2D monolayer material with evenly distributed holes and nitrogen atoms, named C$_{2}$N-$\emph{h}$2D (briefed as C$_{2}$N hereafter), has been synthesized via a bottom-up wet-chemical reaction\cite{C2N-syn}. Because of its tremendous application prospects in electronics\cite{C2N-opt1,C2N-opt2,C2N-opt3,C2N-ele1,C2N-ele2}, gas separation\cite{C2N-gas1,C2N-gas2,C2N-gas3}, and catalysis\cite{C2N-cat1,C2N-cat2,C2N-cat3,C2N-bat1,C2N-bat2,C2N-bat3,C2N-bat5}, C$_{2}$N has drawn much attention from both theoretical and experimental researchers. It is generally believed that monolayer C$_{2}$N is just an ordinary semiconductor with an experimental band gap of 1.9 eV. In this work, using density functional theory (DFT) and TB model, we theoretically predict that the C$_{2}$N monolayer actually is a realistic example of 2D SOTI with gapped edge states and topological corner states. The most desirable aspects of C$_2$N are its large bulk and edge state gaps, which will facilitate the detection of the corner states in future experiments. Furthermore, we demonstrate that the topological corner states are robust against disorder and C$_6$ symmetry breaking of zero dimensional C$_2$N. Similar nontrivial second-order topological phase has also been found in other three C$_2$N-like 2D monolayers and one of them was synthesized experimentally. Considering two of them have been synthesized in the experiments, C$_{2}$N and its analogues are excellent platforms for studying the excellent properties of SOTIs. \\
\section*{METHODS}
\indent Our first-principles calculations were carried out in the framework of generalized gradient approximation (GGA) for exchange-correlation potential in the form of Perdew-Burke-Ernzerhof (PBE)\cite{PBE} implemented in Vienna ab initio simulation package (VASP)\cite{vasp1,vasp2}. A vacuum space of 15 {\AA} was used to avoid couplings between two neighboring slabs. The first Brillouin zone was represented by the Monkhorst-Pack special k-point scheme of 7$\times$7$\times$1 grid meshes. The energy cutoff, convergence criteria for energy, and residual force were set to be 500 eV, 10$^{-6}$ eV, and 0.01 eV/{\AA}, respectively. The bulk band structures were calculated using HSE06\cite{HSE}. The edge states of the semi-infinite structure were obtained with the tight-binding method and maximally localized Wannier functions (MLWFs)\cite{MLWFs1,MLWFs2} constructed with the wannier90 code.\\
\section*{RESULTS}
\begin{figure}
\includegraphics[trim={0.0in 0.0in 0.0in 0.0in},clip,width=5.4in]{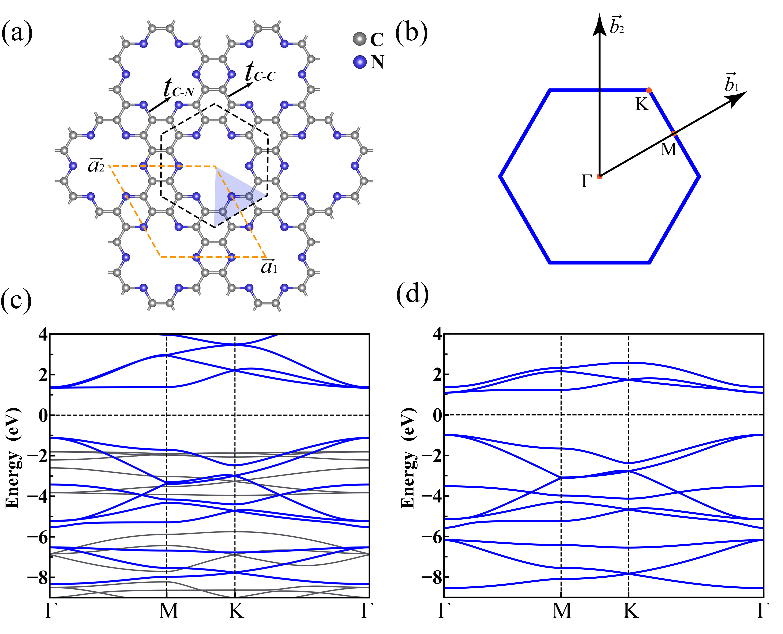}\\
\caption{(a) Crystal structure of C$_{2}$N. The grey and blue spheres represent C and N atoms, respectively. The orange and black dotted lines, respectively, represent normal primitive and Wigner-Seitz unit cell of the monolayer; (b) The 2D Brillouin zone of C$_{2}$N; (c) The bulk electronic band structure of C$_{2}$N obtained by HSE06, blue and gray bands represent the subspaces originated from p$_z$ and s+p$_x$+p$_y$ orbitals, respectively; (d) The band structure of C$_{2}$N obtained from TB model.}\label{fig1}
\end{figure}

\indent The 2D crystal structure of C$_{2}$N monolayer is illustrated in Fig. \ref{fig1}(a). It is a typical 2D porous material and can be structurally obtained from a $2\sqrt{3}\times 2\sqrt{3}$ graphene supercell by removing a carbon six ring and then replacing the C atoms on the edge of the resulted holes with N atoms. The space group of C$_2$N is P6/mmm (No. 191) and it can be generated by the rotation operations C$_{3z}$, C$_{2z}$ and mirror operations M$_y$, M$_z$. The lattice constants optimized by the first-principles calculations are a = b = 8.333 \AA, in good agreement with previous experiments\cite{C2N-syn} and calculations\cite{C2N-opt2,C2N-gas2,C2N-gas3}. Considering the negligible spin-orbit coupling of C and N atoms, C$_{2}$N can be effectively treated as a spinless system. The bulk electronic band structures of C$_{2}$N obtained by HSE06 and PBE are presented in Fig. \ref{fig1}(c) and Fig. S1(b) in the supplemental materials\cite{SM}, respectively. Since the different symmetry between p$_z$ and s plus p$_{x,y}$ orbitals, the bands derived from them form two independent subspaces, as shown in Fig. \ref{fig1}(c). Moreover, a global band gap of 2.45 eV is formed between the valence band maximum (VBM) and conduction band minimum (CBM) at $\Gamma$ point. The VBM and CBM mainly originate from the p$_z$ orbitals of both C and N atoms.\\
\indent Firstly, we examine the topological properties of bulk C$_2$N. Because of the coexistence of the time-reversal and spatial inversion symmetries, the Chern number of the 2D system must be zero. Moreover, the Z$_2$ is always trivial for spinless systems. Except the two topological indexes, \emph{w$_2$} given by 2D winding number of the sewing matrix G(\emph{k}) (modulo 2)\cite{topo-index}, can also be taken as a topological invariant to characterize a SOTI. In space group P6/mmm, the inversion operation \emph{I} can be considered as the product of C$_{2z}$ and M$_z$. Because the eigenvalues of M$_z$ for the p$_z$ states are odd, the eigenvalues of C$_{2z}$ and \emph{I} differ by a sign. Therefore, the eigenvalues of C$_{2z}$ can also be used to calculate the topological index. For 2D hexagonal materials with mirror symmetry $\sigma_z$, the \emph{w$_2$} can be defined as\cite{topo-index,GDY-2}\\
\begin{equation}
(-1)^{w_{2}}=\prod_{i=1}^{4}(-1)^{\left[N_{\mathrm{occ}}^{-}\left(\Gamma_{i}\right) / 2\right]}
\end{equation}
where N$_{occ}^{-}$($\Gamma$) is the number of occupied bands with negative C$_2$ eigenvalues at the TRIM $\Gamma_i$. Under the premise of ignoring spin, there are 39 occupied states in C$_2$N, and the numbers of states with the negative C$_2$ eigenvalues for $\Gamma$ and M points are 21 and 19, respectively. C$_2$N is easily confirmed to be a SOTI with Z$_2$=0 and \emph{w$_2$}=1. It is worth noting that \emph{w$_2$}=1 can be also obtained by only considering the p$_z$ states (the s, p$_x$, and p$_y$ states can only get {w$_2$}=0) implying that the p$_z$ state can not only well-describe its low-energy physics features but also its higher-order topology. The results obtained by TB model(as listed in Tab. S1-2) also supports this argument. To simplify the analysis about the topological properties, we build a TB model with the basis of p$_z$ orbitals as follows: \\
\begin{equation}
\hat{H}=\sum_{i} \epsilon_i+\sum_{i j}\left(t_{i j} \hat{c_{i}}^{\dag} \hat{c_{j}}+h . c .\right)
\end{equation}
In the model, only on-site energies ($\epsilon$$_C$ and $\epsilon$$_N$) and nearest-neighbor hopping between the p$_z$ orbitals ($\emph{t}_{C-C}$ and $\emph{t}_{C-N}$) are considered. By fitting the TB bands to the occupied DFT bands, we obtain the TB parameters $\epsilon$$_C$ = 0.78 eV, $\epsilon$$_N$ = -1.60 eV, $\emph{t}_{C-C}$ = -3.20 eV and $\emph{t}_{C-N}$ = -3.21 eV. As shown in Fig. \ref{fig1}(d), the occupied TB bands are very close to the DFT bands derived from p$_z$ orbitals. Furthermore, the irreducible representations (irreps) of the Bloch states at high symmetry points obtained from first-principles method and TB model match well with each other, as listed in Tab. S1. Namely, the TB model can be effectively used to investigate the electronic topological properties of C$_2$N. The irreps of the Bloch states at the high symmetry points of $\Gamma$, K, and M are closely related to the nontrivial second-order topological phase of C$_2$N, the details will be discussed below.\\
\begin{figure}
\includegraphics[trim={0.0in 0.0in 0.0in 0.0in},clip,width=5.4in]{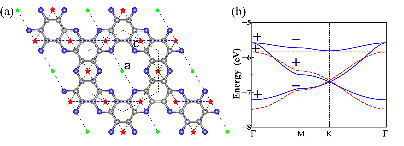}\\
\caption{(a) The sketch of Wyckoff positions of \textbf{1a} (green dot) and \textbf{3c} (red pentagram); (b)The DFT band structure of p$_z$ orbital (blue solid line) and TB band structure (red dashed line) from Wyckoff position 3c ($\epsilon$=-6.39 eV, t$_{pp{\pi}}$=-0.27 eV).}\label{fig2}
\end{figure}
\indent The nontrivial topology of C$_2$N can also be understood from the symmetry properties and Wannier centers of the occupied p$_z$ states. For a one-atom-thick 2D system, the entire two-dimensional plane is mirror-invariant, which means that all states in the plane can be classified into two groups: odd and even under the mirror operation M$_z$. The p$_z$ states are odd, whereas the s, p$_x$, and p$_y$ states are even. In other words, all p$_z$ states form an independent subspace and the in-plane mirror symmetry M$_z$ can be utilized to differentiate p$_z$ states from s, p$_x$, and p$_y$ states. Therefore, it is sufficient to use the subgroup P6mm (No. 183) to investigate the p$_z$ states, and this subgroup can be generated by the operations C$_{3z}$, C$_{2z}$, and M$_y$. There is a C$_6$ rotation operation along z-axis. All Wyckoff positions of the space group are listed in Tab. S3. The C and N atoms of 2D C$_2$N locate at the Wyckoff positions of \textbf{12f} and \textbf{6d} with the local point group of E and C$_{s}$, respectively. The induced irreps from the $p_z$ orbitals at the two positions can be obtained with the theory of elementary band representations (EBRs)\cite{EBR1,EBR2,EBR3}, and the results are listed in Tab. S4 and Tab. S5. However, by summing the number of the valence electrons of C and N atoms, only half of the bands are occupied, which makes the centers of the Wannier functions of the occupied bands may not coincide with the positions of the ions. For the systems with C$_6$ rotation symmetry, the Wyckoff positions of \textbf{1a}, \textbf{2b}, and \textbf{3c} are the maximal, and we only consider the irreps of the wave functions at the high symmetry points of $\Gamma$, K, and M induced from them. The \textbf{1a} and \textbf{3c} Wyckoff positions are labeled with green dots and red pentagrams in Fig. \ref{fig2}(a), respectively. The EBRs induced by the Wannier functions on these Wyckoff positions can be found in the Bilbao Crystallographic Server\cite{bilbao1,EBR2}. Using the formula in the section II in the supplemental material\cite{SM,EBRs}, we establish a series of linear equations taken the EBRs at $\Gamma$, K, and M as the basis set and all the irreps of valence bands of C$_2$N as the solutions. A set of coefficients represents a collection of possible EBRs for the valence bands. Certainly, the coefficients of these equations must be integers. After solving these equations, we discover that the nine valence bands can be induced by the sum of the Wannier functions of maximal Wyckoff positions \textbf{3c}. The Wannier functions at \textbf{3c} positions are A1$^{3c}$, A2$^{3c}$, and B1$^{3c}$. The irreps of the valence bands at the high symmetry points of  $\Gamma$, K, and M are listed in Tab. S1. The mappings between them are
\begin{equation}
\begin{aligned}
A 1^{3 \mathrm{c}} & \mapsto \Gamma_{1}+\Gamma_{5}+M_{1}+M_{3}+M_{4}+K_{1}+K_{3} \\
A 2^{3 \mathrm{c}} & \mapsto \Gamma_{2}+\Gamma_{5}+M_{2}+M_{3}+M_{4}+K_{2}+K_{3} \\
B 1^{3 \mathrm{c}} & \mapsto \Gamma_{3}+\Gamma_{6}+M_{1}+M_{2}+M_{3}+K_{1}+K_{3} \\
\end{aligned}
\end{equation}
where A1 induces the three bands with the lowest energies, whereas the other six occupied bands with higher energies come from A2 and B1. The EBRs of the occupied bands indicate that the centers of Wannier functions locate at the Wyckoff positions of \textbf{3c} that clearly deviate from the ion positions. According to their eigenvalues of symmetry operation, we can obtain the invariant indexes of the three groups of occupied states\cite{hoti-8}. Each index forms a free Abelian additive structure, the non-zero summation of the indexes can be taken as non-trivial criterion of the system. According to the eigenvalues of the C$_2$ at M and the C$_3$ at K, the indexes are
\begin{equation}\label{equ1}
\chi^{(6)}(A1) = (-2, 0);\  \chi^{(6)}(A2) = (-2, 0);\ \chi^{(6)}(B1) = ( 2, 0).
\end{equation}
According to the theory of the index, the above three indexes actually describe the same topological phase and they correspond to the same primitive generator h$_{3c}^6$\cite{hoti-8}. Therefore, the p$_z$ occupied bands of C$_2$N can be seen as $h_{3c}^6 \oplus h_{3c}^6 \oplus h_{3c}^6$ . If the nine bands are separated into the lowest three ones and the other six ones as shown in Fig. \ref{fig1}(d) , the nontrivial topological phase of C$_2$N comes from the lowest three bands because the index of other six bands is equal to (0, 0). If we only consider the A1 orbitals at the 3c Wyckoff positions, the configuration of the A1 orbitals is similar to a Kagome lattice, and the three bands with the lowest energies for C$_2$N are similar to the occupied bands of Kagome lattice\cite{kagome}. The topological invariant \emph{w$_2$} of the lowest three bands is 1. Moreover, the distribution of the C$_2$ eigenvalues is the same as the occupied bands of the Kekul$\acute{e}$-like hexagonal lattice, thus the C$_2$N and the Kekul$\acute{e}$-like hexagonal lattice\cite{kekule1,kekule2} share the same second-order topological nontrivial properties.\\
\begin{figure}
\includegraphics[trim={0.0in 0.0in 0.0in 0.0in},clip,width=5.2in]{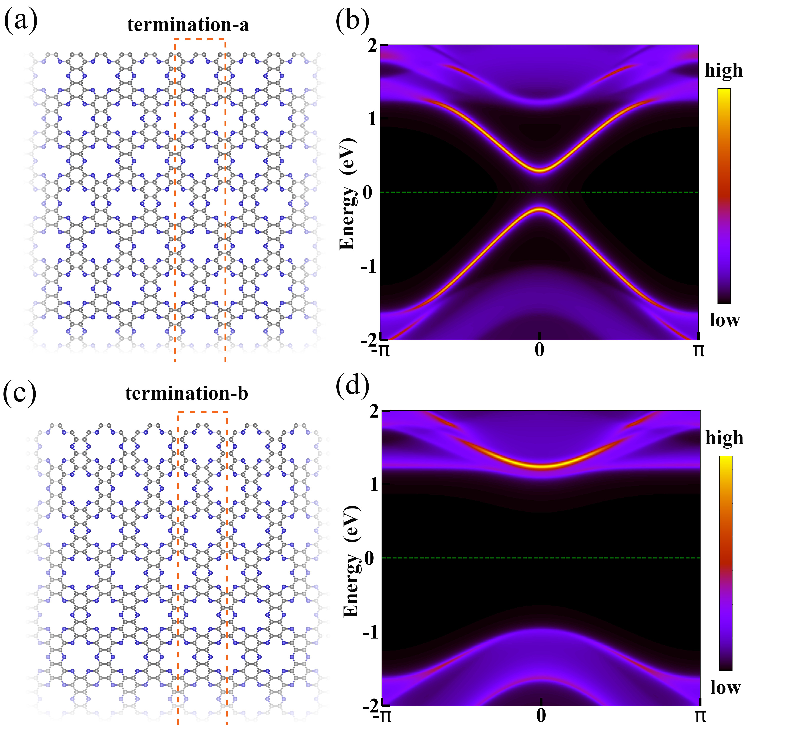}\\
\caption{Semi-infinite C$_{2}$N structures of termination-a edge (a) and termination-b edge (c). The projected edge density of states for termination-a edge (b) and termination-b edge (d), respectively.}\label{fig3}
\end{figure}
\indent To study the edge states of C$_2$N, we cut it into two semi-infinite structures named as termination-a and -b, respectively, as shown in Fig. \ref{fig3}(a) and (c). With the help of the TB Hamiltonian, their edge density of states are calculated and the results are presented in Fig. \ref{fig3}(b) and (d). The results indicate that, between the projected valence bands and conduction bands, there are two gapped edge states for termination-a, whereas there is none for termination-b. The termination dependent edge states can be explained by the filling anomaly\cite{hoti-8} in C$_2$N. As discussed above, the occupied states of C$_2$N can only be induced from the orbitals at the Wyckoff position \textbf{3c} that clearly do not match with the atomic positions of C$_2$N. For the case of termination-a, the \textbf{3c} positions just locate on the boundary, which will produce a kind of charge imbalance ($\rho$). The value of the charge imbalance can be obtained based on obstructed atomic limit approximation\cite{hoti-2} as $\rho = \frac{\#ions - \#electrons}{2}|e|$, where \#ions and \#electrons are the number of ions and electrons per unit cell, respectively. The value of the $\rho$ is proportional to the number of the Wannier centers on one edge per unit cell. If we assume the number of unit cell along the periodic direction is N$_x$, the \#ions and \#elec are 4N$_x$ and 2N$_x$, respectively, for the termination-a edge, and 2N$_x$ and 2N$_x$, respectively, for termination-b. After reducing the number of primitive cells N$_x$, we can acquire $\rho = 2 |e|$  and $\rho = 0 |e|$ for the termination-a and the termination-b, respectively. The value of $\rho/ |e|$ is rightly equal to the number of the edge states. Interestingly, the energy gap of the two edge states, as shown in Fig. \ref{fig3}(b), reaches up to 0.5 eV. Although the energy gap of the two edge states will reduce to 0.32 eV when considering the Wannier Hamiltonian of all orbitals, as shown in Fig. S2, the energy gap of the two edge states will leave enough energy space to accommodate the corner states of C$_2$N that will be discussed below.\\
\begin{figure}
\includegraphics[trim={0.0in 0.0in 0.0in 0.0in},clip,width=5.2in]{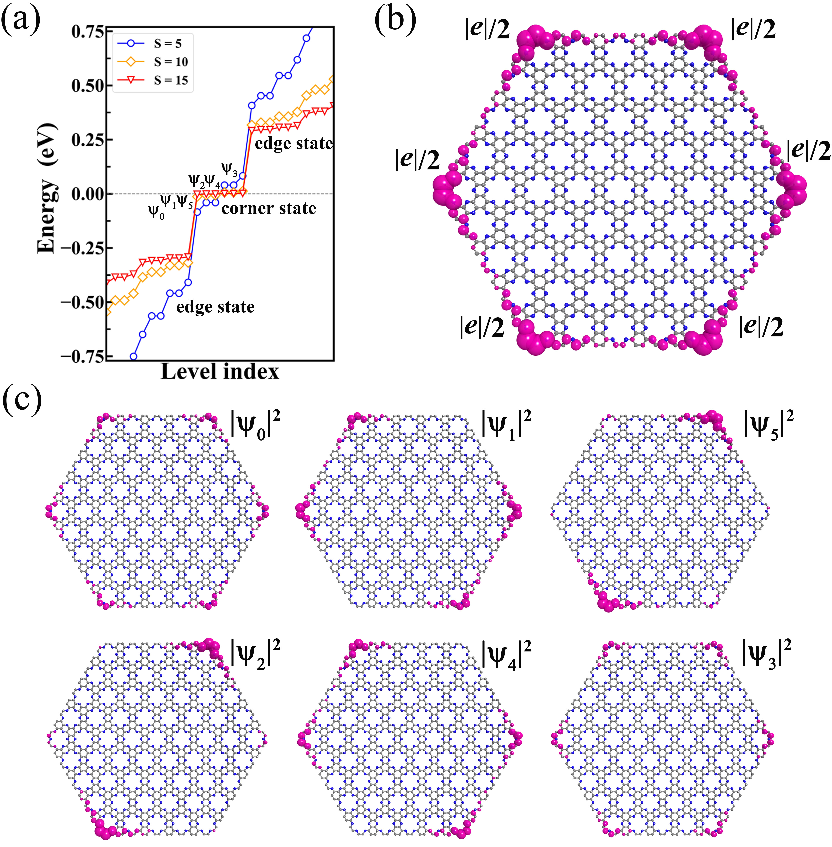}\\
\caption{(a) Energy spectrum of the hexagonal-shaped C$_{2}$N cluster with different size. S is the number of holes (circles surrounding \textbf{1a} position) in C$_{2}$N cluster. (b) The total and (c) the individual real-space wave function distributions of six in-gap topological corner states, The radii of the spheres on each atom denotes the absolute values of square modulus of wave functions.}\label{fig4}
\end{figure}
\indent Besides the gapped edge states, the existence of corner states is another strong evidence for a SOTI. According to the theory of the filling anomaly, the corner charge can be obtained from the symmetry properties of the occupied bulk bands\cite{hoti-8}:
\begin{equation}
Q_{\text {corner }}^{(6)} = \frac{e}{4}\left[M_{1}^{(2)}\right]+\frac{e}{6}\left[K_{1}^{(3)}\right] \bmod e \\
\end{equation}
where $[M_{1}^{(2)}]$ ($[K_{1}^{(3)}]$) is the difference of the number of the states whose C$_2$ (C$_3$) eigenvalue equal to 1 between $\Gamma$ and M (K) points. We find that the terms $[K_{1}^{(3)}]$ and $[M_{1}^{(2)}]$ are equal to 0 and 2, respectively, for C$_2$N, resulting in $Q_{\text {corner }}^{(6)} = 1/2 |e|$. To reveal the distributions of the charges, C$_{2}$N is cut into a hexagonal cluster with armchair edges, and its discrete energy levels are shown in Fig. \ref{fig4}(a). Remarkably, there are six in-gap states ($\psi_{0}$-$\psi_{5}$ as shown in the inset of Fig. \ref{fig4}(a)) around the Fermi energy. Note that, the roughly symmetric distribution of the edge and corner states around the Fermi level results from an approximate chiral symmetry of the edge structure of C$_2$N (the details can be found in section IV in the Supplemental Material\cite{SM}). The results also indicate that $\psi_{1}$ ($\psi_{2}$) and $\psi_{5}$ ($\psi_{4}$) are degenerate. The real space configuration of all the in-gap states can be visualized by their wave function distributions as shown in Fig. \ref{fig4}(b). The in-gap states are well localized at the six corners. In Fig. \ref{fig4}(c), we also show the real-space wave function distributions of the six in-gap corner states separately. Moreover, we find that the density distributions of the wave functions corresponding to each corner mode are exponentially localized around the corners, the results can be found in Fig. S4 in the supplementary materials\cite{SM}. The distributions of the corner states can be well explained by the symmetry of the cluster. The point group of the hexagonal cluster is D$_6$. Thus, we can use the irreps of D$_6$ to analyze the characteristics of the eigenvalues and eigenvectors of the six in-gap states. According to D$_6$ point group, the in-gap states can be written as the linear combination of the isolated p$_z$ states at the six corner as follows:
\begin{equation}
\begin{aligned}
\psi_{m}&=a+\epsilon{_m}b+\epsilon{^2_m}c+\epsilon{^3_m}d+\epsilon{^4_m}e+\epsilon{^5_m}f \\
\epsilon{_m}&=\mathrm{e}^{m\pi\mathrm{\emph{i}}/6} \qquad  m\in{(0,1,2,3,4,5)} \\
\end{aligned}
\end{equation}
where \emph{a}-\emph{f} denote the isolated p$_z$ states at the six corners. According to the theory of equivalence transformation\cite{group}, the equivalence representation of the p$_z$ states $\Gamma^{a.c.}$ can be written as the direct sum of $\Gamma_1$$\oplus$$\Gamma_3$$\oplus$$\Gamma_5$$\oplus$$\Gamma_6$. $\psi_{0}$ will remain unchanged under all the D$_6$ symmetry operations, that the irrep of $\psi_{0}$ is $\Gamma_1$. Analogous to the simple hypothetical SH6 molecule\cite{group}, $\psi_{0}$ state should be non-degenerate and its energy should be the lowest one among the six in-gap states. The irrep of $\psi_{3}$ is $\Gamma_3$ with anti-symmetric characteristics. It should be non-degenerate and its energy is the highest one among the six in-gap states. $\psi_{1}$ and $\psi_{5}$ states conjugate with each other and the irrep of the combination of the two states is $\Gamma_6$. Thus the two states should be degenerate. Similarly, $\psi_{2}$ and $\psi_{4}$ are also degenerate because they are complex conjugate with each other. And the irrep of the two states is $\Gamma_5$. It is worth noting that the splitting of the energy levels of the in-gap states derives from the couplings between the corner states. The splitting will reduce with the increase in the cluster size. Namely, the corner states will become gapless in the limit of large size clusters, the results can be found in Fig. \ref{fig4}(a).\\
\begin{figure}
\includegraphics[trim={0.0in 0.0in 0.0in 0.0in},clip,width=5.4in]{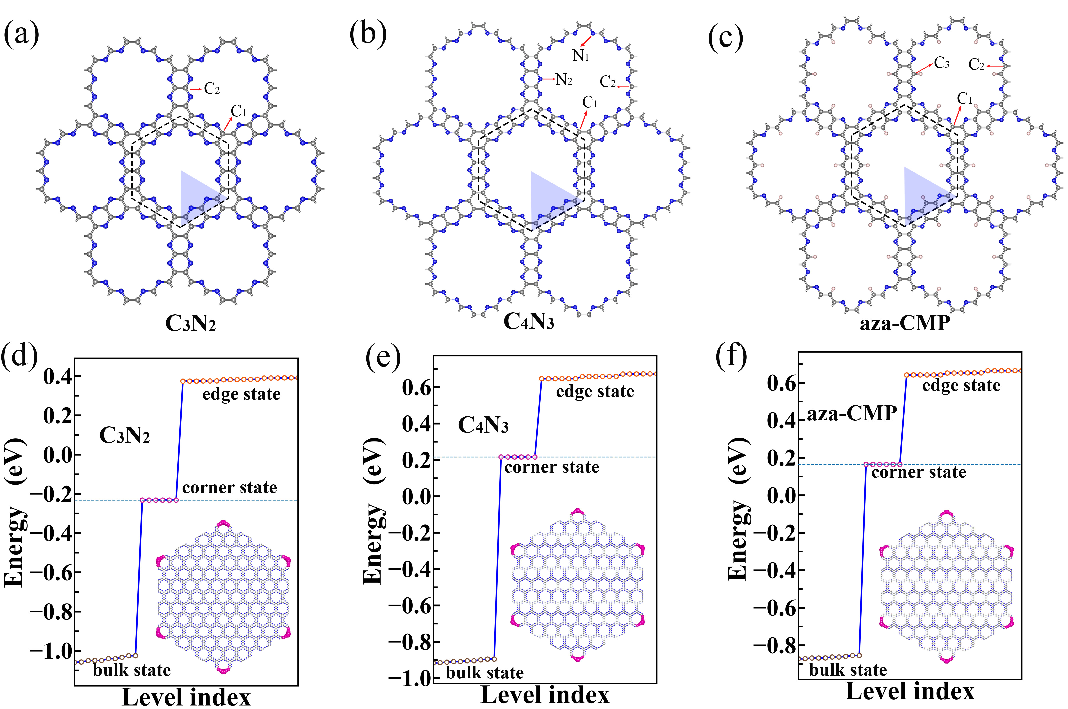}\\
\caption{Crystal structures of C$_3$N$_2$ (a), C$_4$N$_3$ (b), and aza-CMP (c). The blue, grey, and pink spheres represent N, C, and H atoms, respectively. (d)-(f) are the corresponding energy spectra of the hexagonal-shaped clusters, and real-space wave function distributions of six in-gap topological corner states, the radii of spheres on each atom denote the absolute values of the square modulus of wave functions.}\label{fig5}
\end{figure}
\indent For a concrete C$_2$N cluster, the influence of relaxation of the chemical bonds for the boundary atoms and external atom absorptions on its electronic properties should be considered. Therefore, the robustness of the corner states against external perturbations is investigated by adding a random on-site energy to the edge atoms in TB model, and the results can be found in Fig. S5. Despite the intensity of the disorder (the difference between highest and lowest values of random on-site energies) reaches up to 1.0 eV that is much larger than the gap of topological edge states, the eigenvalues of the corner states still remain in the gap of the edge states. However, the degeneracy of the in-gap states are destroyed. Moreover, although the total real-space wave function distributions of all the corner states is very similar to the case without disorders, its distribution symmetry is broken, as shown in Fig. S5(c). It means that strict symmetries of the cluster are not necessary for the appearance of topological corner states. We further calculate the spectrum of the hexagonal cluster introduced a termination-b disorder in one edge and a termination-b edge, respectively, to study the impact of termination-b disorder and termination-b edge on the corner states, as shown in Fig. S6 and Fig. S7. Although two defect states appear between the energies of corner and edge states, the localized corner states remain when a termination-b disorder is introduced in one edge of the hexagonal cluster, as shown in Fig. S6. However, when one edge of the hexagonal cluster is replaced with a termination-b edge, the number of corner states around the Fermi level decreases from six to four, as shown in Fig. S7. The remained four corner states still locate around the Fermi level. Namely, the localized corner states is robust to termination-b disorders, however, the number of the localized corner states will depend on the number of the termination-b edges. Introduction of disorders or weak symmetry-breaking perturbations is analogous to give the symmetric Hamiltonian a continuous transformation. And the transformed Hamiltonian belongs to the same topological classification as the original symmetric one as long as the bulk and edge gaps are preserved. In other words, as long as the bulk and edge gaps remain, a cluster with a perturbation still has equivalent corner states to its original system. Namely, the corner states are still protected by the original bulk topology with strict symmetries. Moreover, under the protection of the bulk topology with strict symmetries, the SOTI can also be categorized into two groups: “intrinsic” and “extrinsic”.\cite{hoti-5} As for the intrinsic one, the corner states of the SOTI are termination-independent. Whereas for the extrinsic SOTI, its corner states are termination-dependent. The above results indicate that the corner states of C$_2$N are "extrinsic," or termination-dependent ones. Although accurately controlling the morphology and specific edges of C$_2$N are still unexplored area, because C$_2$N and graphene have a similar lattice, it is possible to cut  C$_2$N to specified edges by referring the studies on cutting graphene,   the details can be found in Section VII of Supplemental Materials\cite{SM,sup_1,sup_2}. Once we obtained proper termination, the corner states of C$_2$N cluster is robust against disorders of the edges, which is favorable for experimental observation.\\
\indent The Wigner-Seitz unit cell of C$_2$N as shown in Fig. \ref{fig1}(a) indicates that the C$_2$N monolayer can be obtained by applying C$_6$ rotation and translational symmetries to a three-atom ligand (blue shadow area in the figure). Moreover, the ligand can be replace with other odd number C-N atom chains to produce new CN monolayers, such as C$_3$N$_2$ and C$_4$N$_3$ as shown in Fig. \ref{fig5}(a) and (b), respectively. Furthermore, partial and total N atoms in the ligand can be replaced by the C-H atomic group to obtain CNH or CH monolayers. A well-known one with this replacement is the experimentally synthesized aza-fused microporous polymers as shown in Fig. \ref{fig5}(c). We take C$_3$N$_2$, C$_4$N$_3$, and C$_5$N$_2$H as examples to investigate their electronic structures and topological properties. The band structures from first-principles calculations are presented in Fig. S8. The results reveal that these three materials are semiconductors with the band gaps of 2.0 eV, 1.49 eV, and 1.64 eV for C$_3$N$_2$, C$_4$N$_3$, C$_5$N$_2$H, respectively. The topological index \emph{w$_2$} of the occupied states for all the three 2D materials are 1, implying they are SOTIs. With the help of TB models only considering the nearest-neighbor hoppings of p$_z$ orbitals (the detail TB model parameters can be found in Tab. S6), we find that similar corner states appear in the three materials, as shown in Fig. \ref{fig5}(d)-(f).\\
\indent Finally, considering the similar structure between C$_2$N and graphene, C$_2$N also has potential to form Moiré superlattice structures (details can be found in section IX in the Supplemental Material\cite{SM,sup_3,twist_BN,twist_MoS2}). However, its twisted structures will totally differ from those of graphene due to the inequivalence between C and N atoms. Moreover, the flat-bands of twisted bilayer C$_2$N will be similar to those of boron nitride and MoS$_2$\cite{twist_BN,twist_MoS2} rather than that of twisted bilayer graphene due to its semiconductor nature. Then, we can expect that flat-bands should appear around valence band maximum or conduction band minimum of bilayer C$_2$N at specific twist angles as that of twisted boron nitride and MoS$_2$. When the flat-bands appear, the interactions between the flat-bands and the localized edge and corner states will be interesting issue. Although it is beyond the scope of this paper, the issue deserves thorough investigation in the future.\\
\section*{CONCLUSION}
\indent In summary, the nontrivial topology of C$_2$N is revealed by the nonzero bulk topological invariant, gapped edge states, and in-gap topological corner states. Detailed symmetry analysis reveals that the nontrivial topological properties come from the three bands with the lowest energies. Further calculations verify that the corner states of C$_2$N are robust against edge disorders. Moreover, other three C$_2$N-like materials were found to be similar SOTIs, including the experimentally synthesized aza-CMP.\\
\indent The corner states of the four SOTIs proposed in present work can be observed experimentally using spectroscopic approach\cite{con_1} since the energy of the in-gap topological corner states distinguishes from their corresponding bulk and edge states. Moreover, the exponentially localized corner states in real space of the four SOTIs can be investigated with electron energy-loss spectroscopy\cite{con_2}. The novel properties of the corner states, such as resonant tunneling, tunable by external electric or magnetic field\cite{con_3,con_4,con_5}, lasing behavior with low threshold and high spontaneous emission coupling factor\cite{con_6}, and valley-selective corner states\cite{con_7}, will encourage people to investigate the potential of the localized electronic corner states. The results in present work provide realistic candidates for studying the 2D SOTIs in experiments and examining the fascinating corner states of the electronic materials.\\
\begin{acknowledgments}
This work is supported by the National Natural Science Foundation of China (Grant No. 11804287, 11874315, 11574260), Hunan Provincial Natural Science Foundation of China (2019JJ50577, 2021JJ30686) and Scientific Research Fund of Hunan Provincial Education Department (18A051).
\end{acknowledgments}
\bibliographystyle{apsrev4-1}
\bibliography{li}
\end{document}